\documentstyle[aps]{revtex}

\def\be{\begin{equation}}
\def\ee{\end{equation}}
\def\ba{\begin{eqnarray}}
\def\ea{\end{eqnarray}}
\begin{document}
\input epsf
\draft
\twocolumn[\hsize\textwidth\columnwidth\hsize\csname
@twocolumnfalse\endcsname
\preprint{PURD-TH-01-01}
\date{May 2001}
\title{Controlled transfer of quantum amplitude via modulation
of a potential barrier: numerical study in a model of SQUID}
\author{M. Crogan,$^{1,2}$ S. Khlebnikov,$^1$ 
and G. Sadiek$^{1,}$\footnotemark}
\address{
${}^1$Department of Physics, Purdue University, West Lafayette, IN
47907, USA \\
${}^2$School of Electrical and Computer Engineering, Purdue University,
West Lafayette, IN 47907, USA}
\maketitle
\begin{abstract}
We numerically integrate the time-dependent Schr\"{o}dinger equation
in a single-degree-of-freedom model of SQUID with a variable
potential barrier between the basis flux states. We find that linear
superpositions of the basis states, with relatively little residual
excitation,
can be formed by pulsed modulations of the barrier, provided the pulse
duration exceeds the period of small oscillations of the flux. 
Two pulses applied in sequence exhibit strong interference effects, which
we propose to use for an experimental determination of the decoherence
time in SQUIDs.
\end{abstract}
\pacs{PACS: 85.25.Hv, 03.67.Lx \hspace{0.9in} PURD-TH-01-01 \hspace{0.9in}
quant-ph/0105038}
\vskip2pc]
\footnotetext[1]{
On leave of absence from Physics Department, Ain Shams University, 
Cairo, Egypt.}
\addtocounter{footnote}{1}
\section{Introduction}
Ability to manipulate linear superpositions of quantum states
is expected to lead to significant advances in information processing,
the paradigm that became known as quantum computing \cite{review}. 
In addition, if the basis states are macroscopically 
distinct, one may be able to gain useful insights into how
quantum coherence is destroyed (or maintained) in the macroscopic
world \cite{decoh}. 
 
Coherent superpositions of macroscopically distinct states
have been obtained in two recent experiments \cite{suny,delft-mit}
on superconducting
quantum interferometers (SQUIDs). The SQUIDs were in the regime where
the basis states, corresponding to different values of the magnetic flux,
were separated by a potential barrier.
In the first experiment \cite{suny}, a coherent superposition 
was obtained by exciting a SQUID
with a pulse of microwave radiation, thus bringing the system closer
to the top of the potential barrier separating the basis states. 
In the second experiment, 
the barrier was low by design \cite{delft-mit}. 

For quantum information processing, on the other hand, it is essential
that the system can be excited {\em and} subsequently 
deexcited so as to produce a controlled superposition of stable,
low-lying states. The energy of these states, after deexcitation, should
be much lower than the height of the potential barrier separating them,
to ensure that no unwanted transitions occur afterwards. 
The protocols used in refs. \cite{suny,delft-mit} do not allow for such 
deexcitation.

It is clear that a system with a {\em fixed} low-height 
barrier will not be able to satisfy the above requirement. 
Thus, it is natural to look at systems with
{\em variable} potential barriers: lowering the barrier, by some external
means, will correspond to excitation, while restoring it to the original 
height, to deexcitation. The combined effect of the two operations will be
writing a stable quantum superposition to the device.

While at present it is far from obvious that SQUIDs will become 
the underlying technology for quantum computers, these systems do allow
modulation of the barrier height (which in this case is 
the Josephson coupling energy) by an external agent, such as electric
current or magnetic field. Indeed, the height of the barrier was adjustable
in the experiment of ref. \cite{suny}.

The aim of the present work was to find, through numerical integrations,
if modulation of the potential barrier in such a system, by a pulse
of current or field, can be useful
for obtaining controlled superpositions of basis states, and what
the required durations of pulses may be.\footnote{
Our method is different from that recently proposed in ref. \cite{adinv},
where one changes the biasing flux through a SQUID, while keeping
the Josephson energy fixed (and small).}
As a model, we used
the reduced, single-degree-of-freedom, description of a SQUID with the
parameters adopted from ref. \cite{suny}.

Our main result is that rather clean transfers of the quantum amplitude
in such a system are possible within a few tens of picoseconds. (In
Sect. 3 we present a measure of how ``clean'' the transfer is.)
This time scale determines the switching time, $t_0$, of the qubit. The
number of useful operations that the qubit can perform is limited by the
ratio $t_d / t_0$, where $t_d$ is the decoherence time, as well as by another
factor, which we discuss in Sect. 3.

Decoherence \cite{decoh} is an intrinsically many-body effect and 
cannot be studied in the single-degree-of-freedom model. That model, 
however, helps to
identify interference phenomena that can be used for an experimental 
measurement of the decoherence time in SQUIDs. In Sect. 4, we propose to 
use, for that purpose, interference between two consecutive pulses.
This method measures directly the decoherence time, as opposed to
the rate of dissipation, which can be measured by other methods
\cite{dec_time} but then has to be related to $t_d$.

\section{The model}
We start with the usual model of a SQUID as a system with a single degree
of freedom, corresponding to the magnetic flux through the loop.
Fluxes are conveniently measured in units of the flux quantum $\Phi_0$.
If we denote the total flux in these units as $y$, and the biasing flux 
by ${\bar y}$, the Schr\"{o}dinger equation for 
the wave function $\Psi(y, t)$ will read
\be
i\hbar \partial_t \Psi = \left[
- \frac{E_C}{\pi^2} \partial_{y}^2 + E_L (y - {\bar y})^2 - E_J(t) 
\cos(2\pi y)
\right] \Psi \; ,
\label{se1}
\ee
where $E_C$ is the charging energy due to the junction's capacitance $C$:
$E_C = e^2 / 2C$; $E_L$ is the magnetic field energy, due to the
inductance
$L$: $E_L = \Phi_0^2 / 2L$; and $E_J$ is the Josephson coupling energy.
In our case, the latter depends on time, causing a variation in the height
of the potential barrier separating states with different values of the
flux.

In this paper, we only present results for the case when the SQUID is
biased by exactly half of the flux quantum:
\be
{\bar y} = 1/2 \; ,
\label{half}
\ee
although results for an arbitrary bias can be obtained similarly.
In the case (\ref{half}), the potential in eq. (\ref{se1}) is symmetric
about the point $y = 1/2$; in particular, it has two degenerate
minima. Accordingly, we introduce a shifted variable:
\be
x = y - 1/2 \; .
\label{xdef}
\ee
We also define a new time variable $\tau$ via
\be
t = \hbar \tau \; .
\label{tau}
\ee
Frequencies, with respect to this new time, are measured in energy units,
for which we use degrees Kelvin. 
Therefore one unit of time $\tau$, $\Delta \tau = 1$, corresponds to
\be
\Delta t = \frac{\hbar}{k_B\times 1~{\rm K}} = 7.64~{\rm ps}
\label{dt}
\ee
of the physical time. Finally, we shift the potential down by a constant
equal to the unperturbed Josephson energy $E_0$. This corresponds to using
the wave function $\psi=\Psi\exp(iE_0t/\hbar)$. Eq. (\ref{se1}) then leads to
\be
i \partial_{\tau} \psi = \left[
- \frac{E_C}{\pi^2} \partial_x^2 + E_L x^2 + E_J(\tau) \cos(2\pi x) - E_0
\right] \psi \; .
\label{se2}
\ee

\begin{figure}
\leavevmode\epsfxsize=3.25in \epsfbox{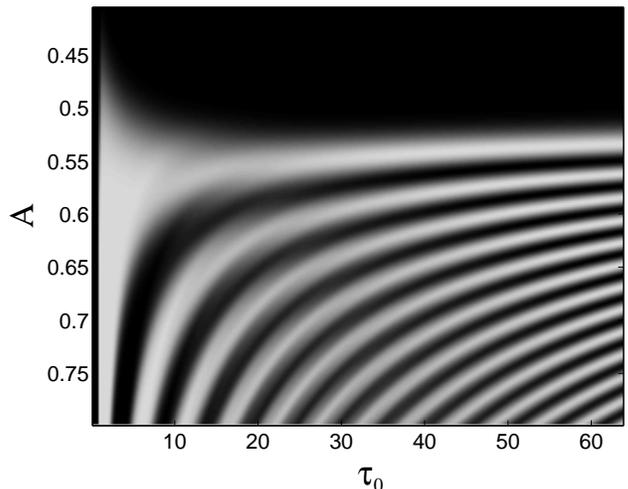}
\caption{Grayscale plot of the probability $P_L$ for the system to remain in
the left well after a Gaussian pulse of magnitude $A$ and duration $\tau_0$.
Darker color means larger $P_L$, i.e. smaller transfer of the
amplitude to the right.}
\label{fig:2dplot}
\end{figure}

We use parameters of the SQUID described in ref. \cite{suny}: $E_C = 9$
mK, $E_L = 645$ K, $E_0 = 76$ K. The time-dependent Josephson energy is taken 
in the form
\be
E_J(\tau) / E_0 = 1 - A \exp[-(\tau - \tau_c)^2 / \tau_0^2] \; .
\label{pulse}
\ee
This corresponds to a Gaussian (in time) pulse of current through the
circuit
that controls the Josephson coupling. If the time $\tau_c$,
corresponding to the center of the pulse,
is noticeably larger than the duration of the pulse, $\tau_0$,
the initial value of $E_J$, $E_J(0)$, is practically indistinguishable from
$E_0$. This initial value corresponds to a potential barrier in
(\ref{se2})
high enough for the tunneling between different flux states to be
inefficient. Accordingly, we can talk about the left and right ground states,
which are the lowest states in each well.
Near $\tau = \tau_c$, however, the barrier is lower, and if
\be
A \geq A_{\rm cr} \equiv 1 - E_L / 2\pi^2 E_0 = 0.57
\label{Acrit}
\ee
it disappears altogether. At large times, the barrier comes back to its
original
height.

\section{Numerical results}
We numerically integrated eq. (\ref{se2}), with a Gaussian pulse of the
form (\ref{pulse}).
In the preparation of each run, we started by relaxing the system to
its true ground state wavefunction (which is a superposition of the
left and right ground states). Then we selected (and normalized) the left
half of that wavefunction as our initial state. After that we numerically
followed the system's evolution.

The magnitude of the pulse, $A$, and the
duration $\tau_0$ were varied, with the goal to achieve either
a complete transfer of the amplitude between the left and right wells, 
or a superposition state. We concentrated primarily on values of $\tau_0$
large enough for the transition to be adiabatic, so that in the end
the system was not significantly excited beyond the left and right ground 
states.

Throughout the
course of the evolution, we monitored a number of quantities. One of these
was the probability $P_L$ to find the system in the left well after the
pulse was completed. It is shown in Fig. \ref{fig:2dplot} 
for a range of values of $A$
and $\tau_0$. For values of $A$ larger than the critical value (\ref{Acrit}),
we observe that $P_L$ is quasiperiodic with respect to $\tau_0$, and the
period decreases as $A$ increases. We interpret this by noting that for
such values of $A$, the potential {\em during} the pulse has a minimum at
$x=0$, about which the system can oscillate. Duration of the pulse will
determine in what phase of these oscillations the system will be deexcited.

Theoretically, we expect oscillations of $P_L$, due to tunneling back
and forth between the two wells, even when $A$ is well
below the critical value. However, the corresponding timescale is too
long for such oscillations to show up in the figure.

Fig. \ref{fig:2dplot} indicates that multiple choices of the parameters
may lead to the same final $P_L$. The presence of
decoherence in any realistic system suggests that one should choose 
the smallest possible $\tau_{0}$,
so as to achieve the maximal number of quantum memory
switches before the device decoheres. On the other hand, a too small
$\tau_0$ will result in overexcitation, and after
many such switches the control of the system will be lost. (One
may contemplate reducing excitation through cooling; however,
any cooling apparatus is likely to become a significant source
of decoherence.) Thus, in addition to the quality factor $Q=t_d/t_0$,
where $t_d$ is the decoherence time, we introduce a ``fidelity factor''
\be
F = (E_b - E_g) / \Delta E \; ,
\label{fid}
\ee
where $E_b$ is the energy at the top of the unperturbed potential 
barrier, $E_g$ is the ground state energy, 
and $\Delta E$ is the increase in energy after one switch.
Decreasing $t_0$ to achieve a larger $Q$ leads, via the time-energy 
uncertainty, to an increase in $\Delta E$ and a smaller $F$, not
unlike how increasing the clock rate in ordinary computers leads to 
overheating.

\begin{figure}
\leavevmode\epsfxsize=3.25in \epsfbox{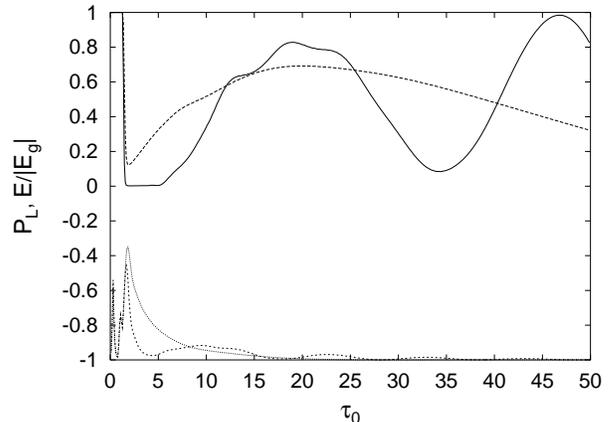}
\caption{Probability $P_L$, as a function
of the pulse duration, for $A=0.59$ (solid line) and $A=0.53$ (long dashes).
The ratio of the final state energy to the magnitude of the ground state 
(initial) energy is shown by short dashes and dots, respectively. This
ratio is a measure of the excitation energy supplied to the system by the pulse.}
\label{fig:energy}
\end{figure}

To study the ``fidelity'' of transitions in more detail, we now focus on
two values of $A$: $A=0.53$ and $0.59$ (corresponding to two sections of
Fig. 1). Both are close to the threshold (\ref{Acrit}). However, the first 
is below that threshold, so tunneling effects are present. In Fig. 
\ref{fig:energy} we show the corresponding probabilities $P_L$ together
with the rescaled final energy $E$ of the system.
In our case, $E_b = 0$ (because of the way the potential has been shifted),
and $E_g$ is negative ($E_g= - 41.1$), so the plotted 
quantity $E/|E_g|$ is simply related to the fidelity parameter (\ref{fid}):
\be
\frac{E}{|E_g|} = \frac{1}{F} - 1 \; .
\label{ratio}
\ee
For example, for $A=0.59$ and $\tau_0 = 5$, we obtain $P_L = 0.004$ (so
the amplitude is almost completely transferred to the right) and
$E = -39.8$, corresponding to $F \sim 30$. 
Increasing $\tau_0$ to $\tau_0 =35$ ($P_L = 0.09$) results in $F\sim 100$.
We also observe that, perhaps contrary to intuition, for short pulses 
the fidelity is higher when the magnitude is slightly above the
threshold (\ref{Acrit}), rather than slightly below.

\begin{figure}
\leavevmode\epsfxsize=3.25in \epsfbox{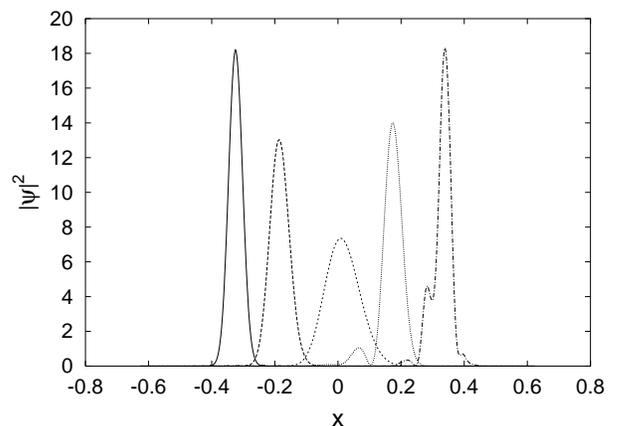}
\caption{A few profiles of the probability density, showing how the
transition occurs for $A=0.59$ and $\tau_0 =5.1$. The maximum of
$|\psi|^2$ moves from left to right as time increases.}
\label{fig:profiles}
\end{figure}

Finally, Fig. \ref{fig:profiles} illustrates how the transition develops
in time. In this case, it is predominantly an overbarrier transition, 
rather than tunneling. In the full many-body theory of the SQUID,
such overbarrier motion of the flux will be associated with dissipation and
decoherence. It remains to see if the time required for
the transition can be made sufficiently short for these effects not to 
present a problem.

\section{Two-pulse interferometry}
We now discuss the behavior of the system under two consecutive
pulses and the possibility to use such two-pulse
sequences for an experimental determination of the decoherence time.
Here we discuss decoherence {\em after} a transition, as opposed
to decoherence {\em during} it, mentioned above.
In its general outline, the method is similar to the two-pulse method
used in studies of magnetic resonance \cite{Ramsey}: in either case,
the final state depends on the phase accumulated by some part of the
wavefunction during the interval between the pulses. 
We now describe our method in more detail.

Suppose that the system is originally in its left ground state, which
we call $\psi_{L0}$. The first pulse will transfer part of the amplitude
to the right well. Because the transition inevitably introduces
some excitation, the state on the right will not be the precise 
right ground state, $\psi_{R0}$, and the state on the left will not 
remain the precise left ground state. 
For simplicity, and during this discussion only,
let us assume that only the admixture of the first excited states,
$\psi_{L1}$
and $\psi_{R1}$, is substantial, while higher excited states can be
neglected.
Then, the state after the first pulse is
\begin{equation}
\psi = c_{L0} \psi_{L0} + c_{L1} \psi_{L1} + c_{R0} \psi_{R0}
+ c_{R1} \psi_{R1} \; .
\label{assum}
\end{equation}
Time evolution of this state is given by
\begin{eqnarray}
\lefteqn{ \psi(t) =  
\left[ c_{L0} \psi_{L0} +  c_{R0} \psi_{R0} + \mbox{} \right. }
\nonumber \\
 & & \left. 
(c_{L1} \psi_{L1} + c_{R1} \psi_{R1}) \exp(-i\omega t) \right]
\exp(-i E_g t/\hbar) \; , \label{evol}
\end{eqnarray}
where $\omega$ is the angular frequency for transitions between 
the ground and the first excited states. (The potential is assumed
left-right symmetric.)

\begin{figure}
\leavevmode\epsfxsize=3.25in \epsfbox{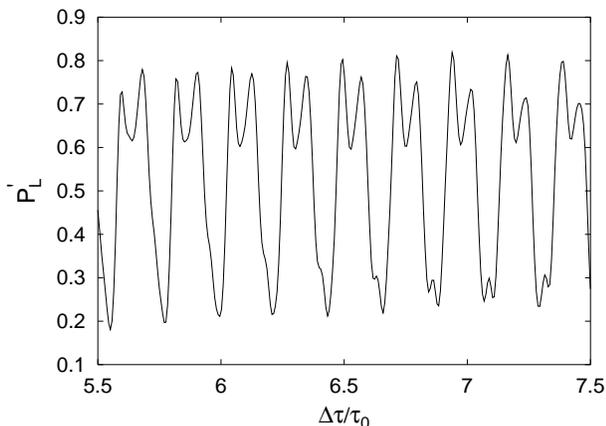}
\caption{Probability that the system remains in the left well after two
consecutive pulses separated by $\Delta \tau$. The pulses were identical
with magnitude $A=0.59$ and duration $\tau_0 = 11.35$.}
\label{fig:sec_pulse}
\end{figure}

If at some time $t$ we apply a second pulse, whose action on the state
(\ref{evol}) is represented by some unitary operator $U$, and then measure
the probability to find the system somewhere in the left well,
the result will be
\[
P'_L(t) = |\psi^*_{L0} U \psi(t)|^2 + |\psi^*_{L1} U \psi(t)|^2 \; ,
\]
which in general is an oscillating function of $t$ with frequency
$\omega$.

The amplitude of such oscillations can be quite
large, as we show in Fig. \ref{fig:sec_pulse}. This
figure is based on numerical integration and does not involve the
simplifying assumption (\ref{assum}).

In practice, the system is coupled to the environment, and this coupling
introduces a certain amount of decoherence. Suppose that evolution of
the environment depends only on which well the system is in. We then
consider two histories of the environment, $\chi_{L}$ and $\chi_{R}$, and
instead of eq. (\ref{evol}) obtain
\begin{eqnarray*}
\psi(t) & = & \left\{ c_{L0} \psi_{L0} \chi_L(t) +  c_{R0} \psi_{R0}
\chi_R(t) \right. \\
 & & + \left. [c_{L1} \psi_{L1} \chi_L(t)+ c_{R1} \psi_{R1} \chi_R(t) ]
\exp(-i\omega t) \right\} \\
 & & \times \exp(-i E_g t/\hbar) \; .
\end{eqnarray*}
The probability to observe the system in the left well after the second
pulse
is now
\[
P'_L(t) = \sum_n \left[
|\psi^*_{L0} \chi^*_n U \psi(t)|^2 + |\psi^*_{L1} \chi^*_n U \psi(t)|^2
\right] \; ,
\]
where $\chi_n$ is a complete system of states of the environment. The
oscillating dependence of $P'_L$ on time is now modulated by the coherence
factor $|\chi^*_R(t) \chi_L(t)|^2$, which decreases in time.
Note that even when this coherence factor becomes essentially zero, the
oscillations
of $P'_L$ remain, due to nonzero matrix elements, 
such as $\psi^*_{L0} U \psi_{L1}$, between the states in the same well.

Thus, due to decoherence, the oscillations of $P'_L$, such as those shown
in Fig. \ref{fig:sec_pulse}, will acquire an envelope. By fitting the
envelope with an exponential
\be
f(\Delta t) = a_1 + a_2 \exp(-\Delta t / t_d) 
\label{exp}
\ee
(or, perhaps, some other function that may emerge on theoretical grounds),
one may be able to extract the decoherence time $t_d$.


\begin{thebibliography}{99}
\bibitem{review} For a review, see A. Ekert and R. Jozsa, 
Rev. Mod. Phys. {\bf 68}, 733 (1996).
\bibitem{decoh} For a review, see W. H. Zurek, Phys. Today
{\bf 44}, 36 (October 1991).
\bibitem{suny}
J. R. Friedman {\it et al.}, Nature, {\bf 406}, 43 (2000).
\bibitem{delft-mit}
C. H. van der Wal {\it et al.}, Science, {\bf 290}, 773 (2000).
\bibitem{adinv}
P. Silvestrini and L. Stodolsky, Phys. Lett. A {\bf 280}, 17 (2001).
\bibitem{dec_time}
C. Cosmelli {\it et al.}, Phys. Rev. Lett. {\bf 82}, 5357 (1999).
\bibitem{Ramsey}
N. F. Ramsey, Phys. Rev. {\bf 100}, 1191 (1955).
\end{thebibliography}
\end{document}